# Disentangling ferroelectric domain wall geometries and pathways in dynamic piezoresponse force microscopy via unsupervised machine learning

Sergei V. Kalinin,[1,*] James J. Steffes[2], Yongtao Liu,[1] Bryan D. Huey,[2] and Maxim Ziatdinov[1,3,†]

[1] Center for Nanophase Materials Sciences, Oak Ridge National Laboratory,
Oak Ridge, TN 37831, USA

[2] University of Connecticut, Materials Science and Engineering,
Storrs, CT 06269-3136, USA

[3] Computational Sciences and Engineering Division, Oak Ridge National Laboratory,
Oak Ridge, TN 37831, USA

Domain switching pathways in ferroelectric materials visualized by dynamic Piezoresponse Force Microscopy (PFM) are explored via variational autoencoder (VAE), which simplifies the elements of the observed domain structure, crucially allowing for rotational invariance, thereby reducing the variability of local polarization distributions to a small number of latent variables. For small sampling window sizes the latent space is degenerate, and variability is observed only in the direction of a single latent variable that can be identified with the presence of domain wall. For larger window sizes, the latent space is 2D, and the disentangled latent variables can be generally interpreted as the degree of switching and complexity of domain structure. Applied to multiple consecutive PFM images acquired while monitoring domain switching, the polarization switching mechanism can thus be visualized in the latent space, providing insight into domain evolution mechanisms and their correlation with the microstructure.

[*] sergei2@ornl.gov
[†] ziatdinovma@ornl.gov

Polarization switching and domain dynamics in ferroelectric materials underpins a broad spectrum of applications, most prominently including non-volatile ferroelectric memories but also materials with giant electromechanical responses, actuators, and many others.[1-5] While capacitive ferroelectric memories remained a niche application for over two decades,[6-8] advances in oxide growth and integration[9] have recently enabled sub-1 V switching and integration of ferroelectrics with magnetic materials, potentially opening the pathway to the low-voltage multiferroic memories.[10] Similarly, domain wall dynamics plays a critical role in other forms of ferroelectric information technology devices, including barrier-based electroresistive systems and domain wall electronics.[11-17]

These considerations necessitate a thorough understanding of domain wall behavior and switching mechanisms on the nanometer scale, including the mechanisms of domain nucleation, motion, and pinning on the topological and structural defects. For decades, this information was available only in special cases via optical microscopy and certain modalities of electron beam probes.[3] However, the development of Piezoresponse force microscopy (PFM) in the mid-90's[18-25] enabled high-resolution, high veracity studies of the polarization and polarization dynamics in ferroelectrics, including imaging, spectroscopy, and dynamics studies on free surfaces and device structures.

From the variety of the PFM imaging and spectroscopic modes, the most direct information on domain dynamics is obtained via dynamic PFM imaging. In this approach, the biased PFM tip is used to scan material surface inducing domain switching. Under certain conditions, switching can be slow compared to the image acquisition time, allowing to visualize the gradual process of ferroelectric domain nucleation and growth[26-29] even for a range of film thicknesses.[30] Similar data can be obtained for the top-electrode devices,[31-34] albeit at typically lower spatial resolution.[35] Finally, PFM imaging in conductive liquids[36-39] that allow to separate low-frequency conductivity enabling switching and high-frequency suppression of ionic dynamics that enables imaging[40,41] can provide similar information.

However, while the observation of bias induced polarization switching is by now routine, the analysis of the data remains a complex issue. Until now, the majority of such analyses included the determination of the overall wall velocity and its correlation with local curvature,[42-45] determination of the statistical properties of the moving and static domain walls, and parameters such as nucleation site densities.[32,46,47] A fundamental bottleneck is the lack of appropriate mathematical tools and workflows for rapid analysis of such data, especially as improved imaging speeds and system stabilities enable increasingly large datasets that become impractical for all but automated analysis. However, the use of simple descriptors such as wall position and velocity limits the scope of possible analyses, whereas more complex descriptors that can also describe local wall geometry typically lead to highly noisy data sets, limiting the extraction of governing physical behaviors. More generally, the wide range of semiquantitative notions phenomenologically developed for describing polarization switching processes lack well defined and rigorous local definitions.

Here, we introduce an approach for the analysis of dynamic PFM data based on variational autoencoder with rotational invariance. This method allows for self-supervised creation of optimal parsimonious (latent) descriptors for the dynamic domain walls, importantly encompassing the presence of multiple rotational variants forming in real materials due to pinning at topological and structural defects, and also robust towards the drift, distortions, and noise. We demonstrate that this approach can establish the basic structural features of the evolving domain structure and use it to describe polarization switching pathways in ferroelectric materials. In addition, we discuss a connection of the inferred latent variables to physical mechanisms.

As a model system, we explore polarization switching in an epitaxial lead zirconate titanate (PZT) thin film comprising 90 degree domains.[48] The 150 nm PZT layer is grown by pulsed laser deposition (PLD) on a $SrTiO_3$ (001) substrate (CrysTec GmbH, Germany), with a heteroepitaxial intermediate conducting oxide electrode (SRO). The deposition is conducted in 100 mTorr oxygen partial pressure at 650 °C, after which the samples are cooled at 1 °C/min to room temperature in a 760 Torr oxygen partial pressure environment.[49] Using the PFM measurements, we explored the domain dynamic as a function of time for constant tip bias, roughly equivalent to the constant electric field. Consecutive PFM images are acquired on the as-received surface during ferroelectric switching from the (001) to the (00-1) states (blue to yellow, respectively, in Fig. 1). This is achieved by continuously applying a DC bias to the conducting probe, which remains in constant contact while raster scanning the ferroelectric thin film. The (applied) local electric field is just sufficient to surpass the ferroelectric coercive field but is effectively only applied for short durations at any given location according to the scanning parameters. Therefore, domain nucleation and growth can be simultaneously excited and directly observed as described in Refs. [29,50]. This sample is especially effective for investigating the efficacy of the variational autoencoder approach because switching proceeds primarily via domain growth upon positive tip biasing and nucleation upon negative biasing.[28] This disparity is attributable to composition and strain gradients through the film thickness, and correspondingly polarity-dependent defect energies and distributions. Specifically, the energetic landscape is relatively uniform when biasing in the positive direction, whereas there are numerous local minima and hence nucleation sites during negative poling.

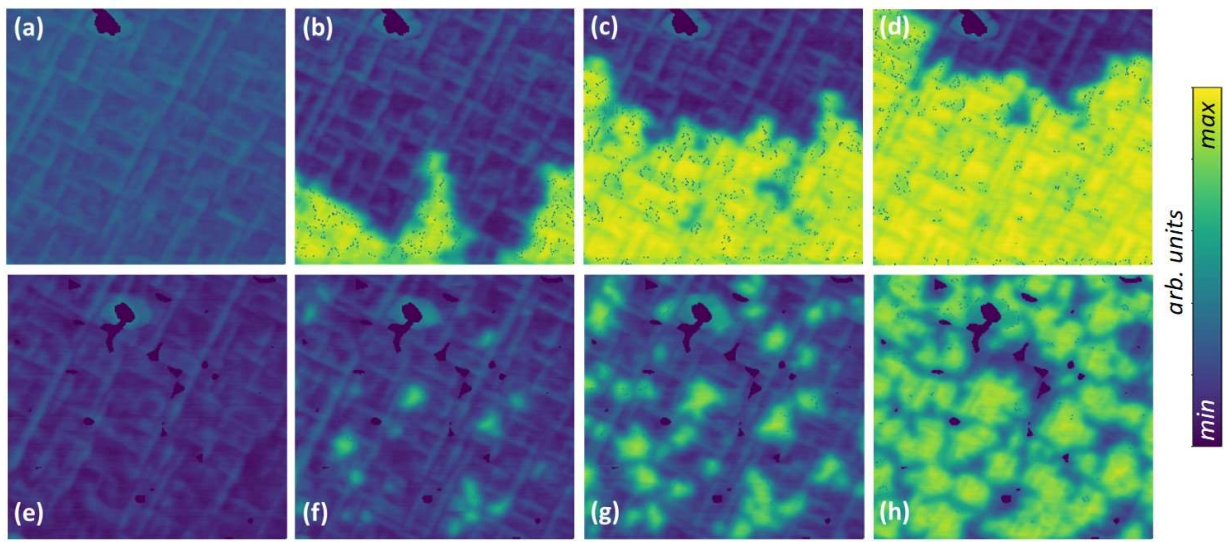

**Figure 1.** The evolution of domain structure in the PZT films during the dynamic PFM experiments. The top row displays frames (a) 0, (b) 20, (c) 30 and (d) 40 during positive switching clearly dominated by domain wall motion; the bottom row presents frames (e) 0 (f) 5 (g) 10 and (h) 20 during negative biasing where domain polarization proceeds principally by nucleation. All images are normalized to (0, 1) range of pixel intensities. The scan sizes are 1.5 µm throughout. The fast and slow scan directions are left-to-right and up-to-bottom, respectively.

The evolution of the domain structure for both scenarios are shown in Figure 1. In the top row, domain nucleation occurs at a small number of sites near the bottom of the imaged region, with the domain wall propagating through the field of view during repeated poling and scanning. Note the strong interactions between the 180° domain wall and preexisting in-plane *a*-domains, which exhibit a pitch on the order of 100 nm with 90° rotations. This gives rise to characteristic domain-front morphologies controlled by a series of local pinning and unpinning events. The nucleation scenario (bottom row), on the other hand, displays dramatically different behavior. Nucleation occurs at multiple sites throughout the film, possibly again with growth mitigated by the pre-existing, cross-hatched, in-plane domain walls, and also new nucleation events continuing to occur as the poling progresses.

This rich spectrum of domain shapes and behaviors observed in Figure 1 brings forth the question of the natural descriptive language for such evolving features. While the domain wall positions could in principle be detected via edge filters, converting such PFM images into effective wall coordinates and thus reducing the data storage and computational requirements, this approach will not capture the full range of observed phenomena and wall morphologies. Furthermore, while the data in Fig. 1 is conveniently already spatially aligned due to initially high speed and stable imaging conditions that minimize lateral drift, PFM and other large microscopy datasets more typically exhibit frame to frame drifts or distortions that are often nonlinear. Here, we confirm the applicability of rotationally invariant variational autoencoders for the computationally efficient and unsupervised analysis of such data.

The general idea of the autoencoder (AE) neural networks is compression of such datasets to a small number of latent variables through a set of convolutional and/or fully-connected layers, and a subsequent reconstruction of the initial image or spectra from this latent representation.[51] The AE training aims to minimize reconstruction loss by optimizing weights of the compression (encoder) and decompression (decoder) parts of network. In this process, the AE finds the optimal parsimonious representation of the original data set. This approach is somewhat similar to classical dimensionality reduction techniques such as principal component analysis (PCA), except that AE can have non-linear layers and hence give rise to more compact latent representations. Variational autoencoders (VAE)[52] share the general encoder-decoder approach of AEs, but utilize a completely different information flow. Here, the decoder and encoder neural networks are used to parametrize the deep latent-variable generative model and the corresponding inference model, respectively. During the training, the encoder outputs parameters of a probabilistic distribution, which is usually chosen to be a diagonal Gaussian. The decoder generates a latent vector by sampling from the encoded distributions (one for each latent dimension) and tries to reconstruct the original object (image or spectra). This VAE set up enforces a continuous, smooth latent space representation allowing the observation of smooth transitions between distinct states. The VAEs can be interpreted as the combination of the AE and Bayesian network concepts.

The application of these machine learning techniques to the data in Fig. 1 will include the creation of a stack of sub-images of predetermined size, similar to the operation of the convolutional neural networks or sliding window transform texture analysis.[53-55] In this process, the movie stack $PR(x, y, t)$ of size $N$x$N$x$M$ is converted into the stack of sub-images $pr_i(x, y)$ of size $n$x$n$, and index arrays $t_i$ and $(x_{0i}, y_{0i})$. In this manner, for each sub-image the original location $(x_0, y_0)$ in the image frame at time $t$ is preserved. The choice of sub-images represents a complicated problem that affects subsequent analysis workflow. For example, analysis of domain wall phenomena can be performed by selecting sub-images centered at the domain walls. However, this approach will necessitate ad-hoc mapping of the domain walls centers, which necessitates human-designed criteria for filter-based or network-based labeling. Secondly, even when centered at the domain wall, it can have arbitrary orientation within the sub-image. It is trivial to show that in this case application of techniques such as AEs, VAEs, along with simpler techniques such as PCA[56-59] or Gaussian Mixture Models (GMMs)[60], will all be limited by the presence of a large number of rotational variants. Effectively, for high quality data and large sub-image sets, the number of variants will be limited by the number of distinguishable (within accessible sampling) rotations of domain walls within the set of sub images. The conventional assumption of continuous translational symmetry (to cope with sub-images that are not necessarily centered at the domain wall) further exacerbates the problem.

To tackle this problem and analyze the general imaging data, here we utilize the rotationally invariant extension of VAE (rVAE).[61] The rVAE represents a special class of VAEs where three of the latent variables are rotation and (optionally) x- and y-offsets, complemented by classical latent variables associated with image content. Thus, rVAE adds rotational and (in this case) offset invariance to the analysis workflow. In other words, it is expected to recognize the images even if they are shifted and rotated with respect to each other. In this case, the sub-images are used to train the rVAE, which compresses the $n \times n$ image down to 3+$m$ variables, including angle, $x$ offset, $y$ offset, and $m$ non-linear latent variables corresponding to image content. In the following we are going to refer to the image content latent variables simply as first latent variable, second latent variable, etc. Each sub-image is parametrized by these variables, and they can be plotted as a function of position $(x_0, y_0)$, visualizing local system behavior. Similarly, changes of latent variables with time provides information into evolutionary pathways of the system. The number of latent variables can be varied depending on the complexity of the problem, i.e. distinguishable features in the data set. Here, the sub-image size can be used to capture domain wall structure at different level of complexity. In general, we note that a good indicator for the choice of number of latent variables is the collapse of the variability in the disentangled lattice-space representation, as will be illustrated below. Both encoder and decoder parts of rVAE were represented by the 2-layer fully connected neural networks, with 512 neurons in each layer activated by a *tanh* function. The training was performed using the Adam optimizer with learning rate of 1e-4 and mini-batches of size 50. The model(s) training and subsequent analysis were performed using a home-built AtomAI package.[62]

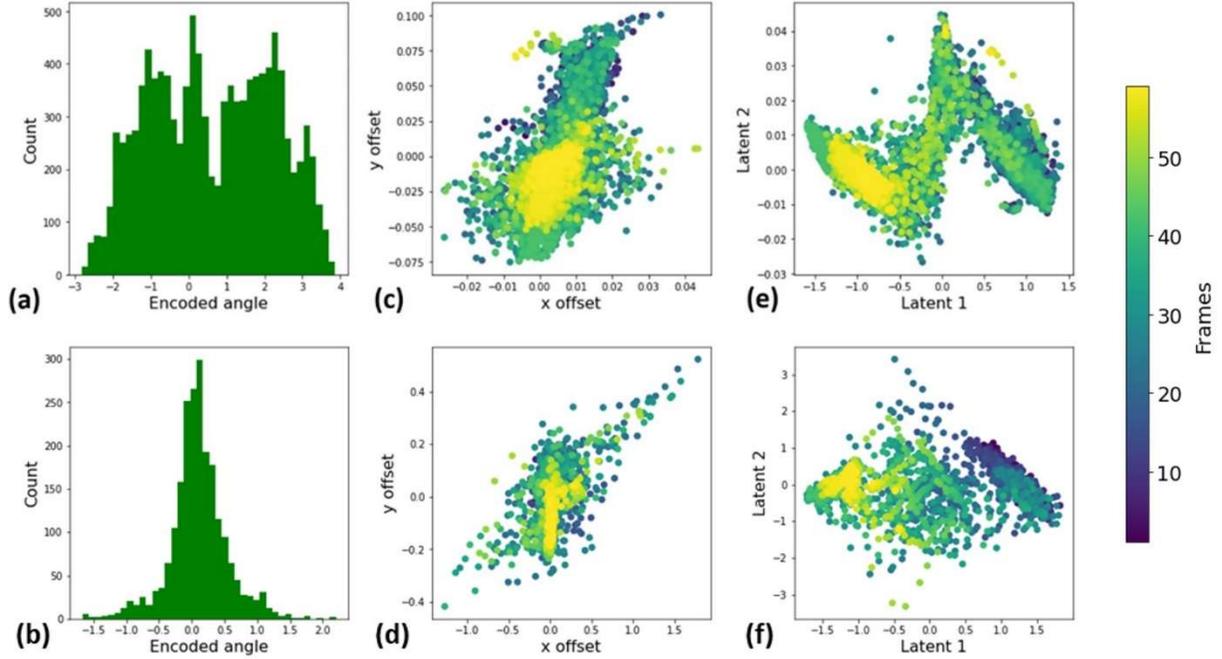

**Figure 2.** (a,b) Latent angle distributions integrated over the full data set, (c,d) offset distributions, and (e,f) latent variable distributions for (a,c,e) $n = 16$ and (b,d,f) $n = 64$ ($n$ is window size for generating sub-images). The color of the point in (c-f) encodes the frame number, and hence color evolution across the latent space reflects the pathway of the process.

To illustrate the rVAE analysis of the dynamic PFM data, Figure 2 shows the distribution of the rotational angle and the distribution of points in the latent space of the system for a small window size ($n = 16$). In this case, the angle distribution has multimodal distribution, corresponding to the preferential orientation of the domain walls along the crystallographic directions. The offset distribution expressed in pixel fractions, Fig. 2 (c), is very narrow along the $x$ and $y$ real-space axes. The calculated latent space distribution, Fig. 2(e), is very narrow in one direction (latent 2, ordinate axis in the plot) and extended in the other (latent 1, abscissa). This behavior suggests that the latent space is essentially 1D in this case, such that the system can be efficiently described by just a single latent variable. In comparison, for a 4x larger window size, the angle distribution becomes centered, Fig. 2(b), and the real-space offset (d) and latent variable distributions (f) are 1 to 2 orders of magnitude broader. This indicates that latent space is now 2D and encapsulates additional information about the wall geometry. The color scale of the points (blue to yellow) is chosen based on the frame number and will be discussed below when analyzing the switching kinetics.

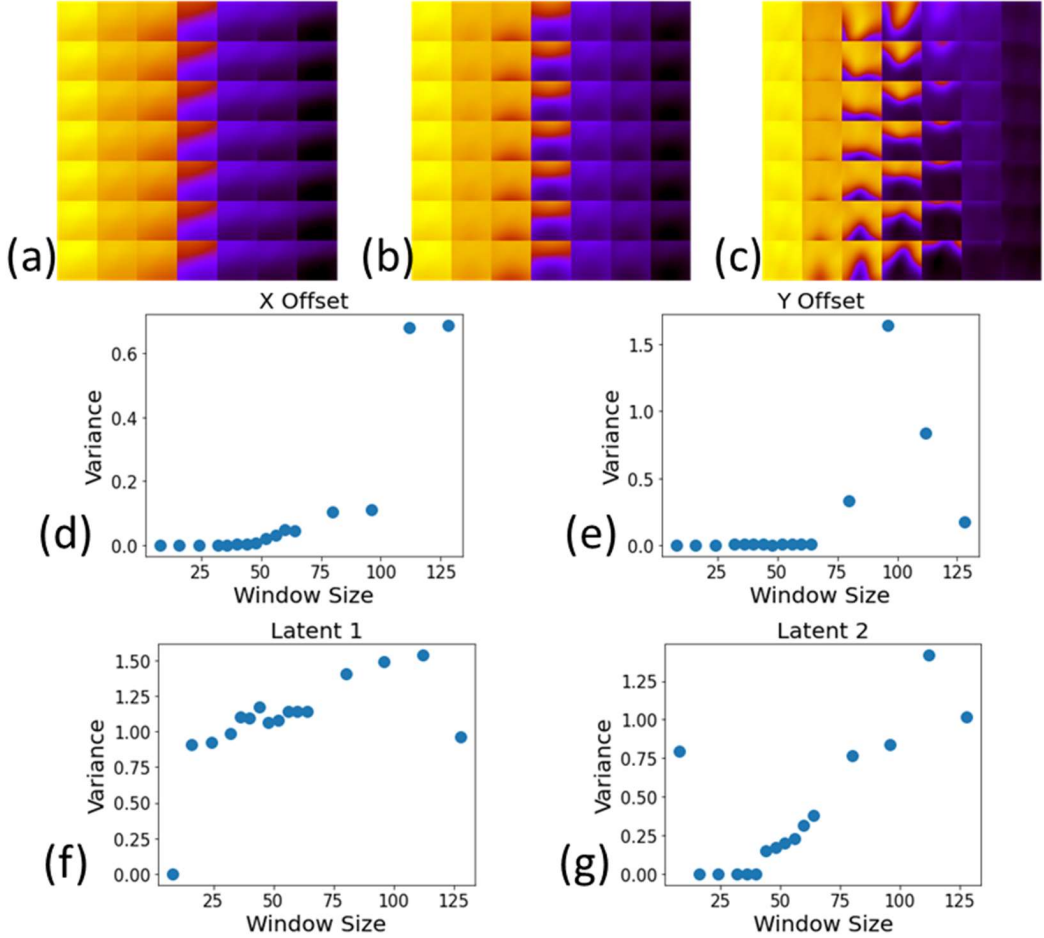

**Figure 3.** (a, b, c) Latent space representations and (d-g) dispersions of x-offset (d), y-offset (e), Latent 1 (f), and Latent 2 (g) as a function of window size ranging from 8 to 128. The dispersion is described by variance. Note that dispersions of latent variables increase with increases in window size.

The natural question is the meaning of the latent variables, i.e. relationship between the coordinates in the latent space and the elements of the domain structure. This relationship is established by the decoder part of the rVAE, in which the set of latent variables is decoded into the sub-images. For the 2D latent spaces, the decoding process can conveniently be represented as shown in Figure 3. Here, an equally spaced grid of points in the 2D latent space is created and decoded into images, reflecting the variability of materials behavior in the latent space. Grids of these images are plotted in (a-c) for increasing window sizes. Note the relationship between the types of representations in Figure 2 (e) and Figure 3 (a). Figure 2 illustrates the distribution of points corresponding to the experimentally observed data in the latent space, while Figure 3 illustrates how the sub-images from the range of points in the latent space would appear.

For small window sizes, the reconstructed sub-images tend to comprise simple geometric shapes, effectively delineating the domain wall. Note that the orientation of the wall is the same across the latent variable space, since it is already captured by the angle latent variable. In this manner, the rVAE was able to capture the characteristic feature of the system and disentangle it into rotation angle and latent descriptors. The variation across the second latent variable (horizontal axis) shows a gradual change of the contrast, indicative of the presence of the continuous variation of the signal in the system containing both 180 and a-c domain walls. Finally, the image variation in the second lateral direction is extremely small, suggesting that the system can effectively be described by a single latent variable. The dispersion of latent variables as a dependence of window size is shown in Figures 3 (d-g). The dispersion of latent variables increases for larger window size, suggesting that the latent variables are more informative. This is because the behavior of domain walls, which is complex during switching, is included for larger window size. Note that the dispersion of the second latent variable is nearly constant and very small (<0.1) when window size is smaller than 40; in these cases, the system can be predominantly described by only one latent variable.

The latent representation for large window size is shown in Figure 3 (c). Here, the latent space is no longer degenerate, and variability in the reconstructed domain geometry is observed in both dimensions, along with the non-negligible variations in the dispersion for both latent variables. The extremely important aspect of this representation stems from the fact that the VAE in general tends to disentangle the characteristic behaviors in the data set. This is the reason they are used extensively in computer vision, image reconstruction, etc. As applied to the image data here, the rVAE clearly disentangles the switching behavior (bottom to top) and domain wall complexity (left to right in Fig. 3c). In this case, this complexity can be roughly associated with domain wall curvature, but more rigorously represents some complex function of wall geometry characteristics for this particular system.

For further insight into this behavior, we establish the relationship between the latent space variables and physics-based descriptors. For all window sizes, the natural descriptor is the degree of switching ($D_s$), which can be determined as

$$D_s = \frac{A_s - A_{us}}{A_s + A_{us}},$$

where $A_s$ is the area of switched region and $A_{us}$ is the area of unswitched region. The switched area and unswitched areas are classified by the median value of latent space, the area (e.g. yellow) with values larger than the median is defined as switched area and the complementary area (e.g. black) is defined as unswitched area. Figures 4a-c show the degree of switching corresponding to latent space with different window sizes. For smaller windows, such as 16 in Figure 4a, the degree of switching only depends on the first latent variation (horizontally), indicating the system can be described by a single latent variable. When window size equals to 32 (Figure 4b), the switch degree also slightly changes along the vertical direction (see the middle column in Figure 4b), suggesting that the second latent variable also reflects the degree of switching. With window size increased to 64 (Figure 4c), changes in the switching degree along the vertical direction become much clearer, indicating that both the first and second latent variable are now related to the degree of switching.

Another physical descriptor is domain wall curvature, which is reflected in the latent space variables for larger window sizes. Here, to determine the domain wall curvature, we first identified the locations of domain walls using the Canny edge filter.[63] Shown in Figure 4d-f are the detected domain walls, which correspond to the latent space manifolds in the insets of Figure 4a-c. Then, the average and the maximum values of the domain wall curvature are calculated for each latent space. Note that there are no detectable domain walls for the fully switched and unswitched parts of the latent space. For the small window size of 16, domain walls only show up in the middle column with an almost identical shape (Figure 4d). For the window size of 32, a relatively small change from the bend-up to bend-down shape in the domain walls curvature is observed in the vertical direction (Figure 4e). This suggests that the second latent variable now also contains information about the domain wall shape. Finally, when the window size increases to 64, there are clear changes in the domain wall curvature in both horizontal and vertical directions (Figure 4f), indicating that both latent variables now describe the domain wall curvature. Overall, a more complex curvature is seen for larger window sizes, indicating that the latent variables inferred by rVAE models trained on larger window sizes are more informative for interpreting the domain walls. The effect of window size on the association between the wall curvature and the rVAE's latent variables becomes even clearer when using a much larger grid, as illustrated in Figure 4g-h for the average wall curvature on an 80*80 grid.

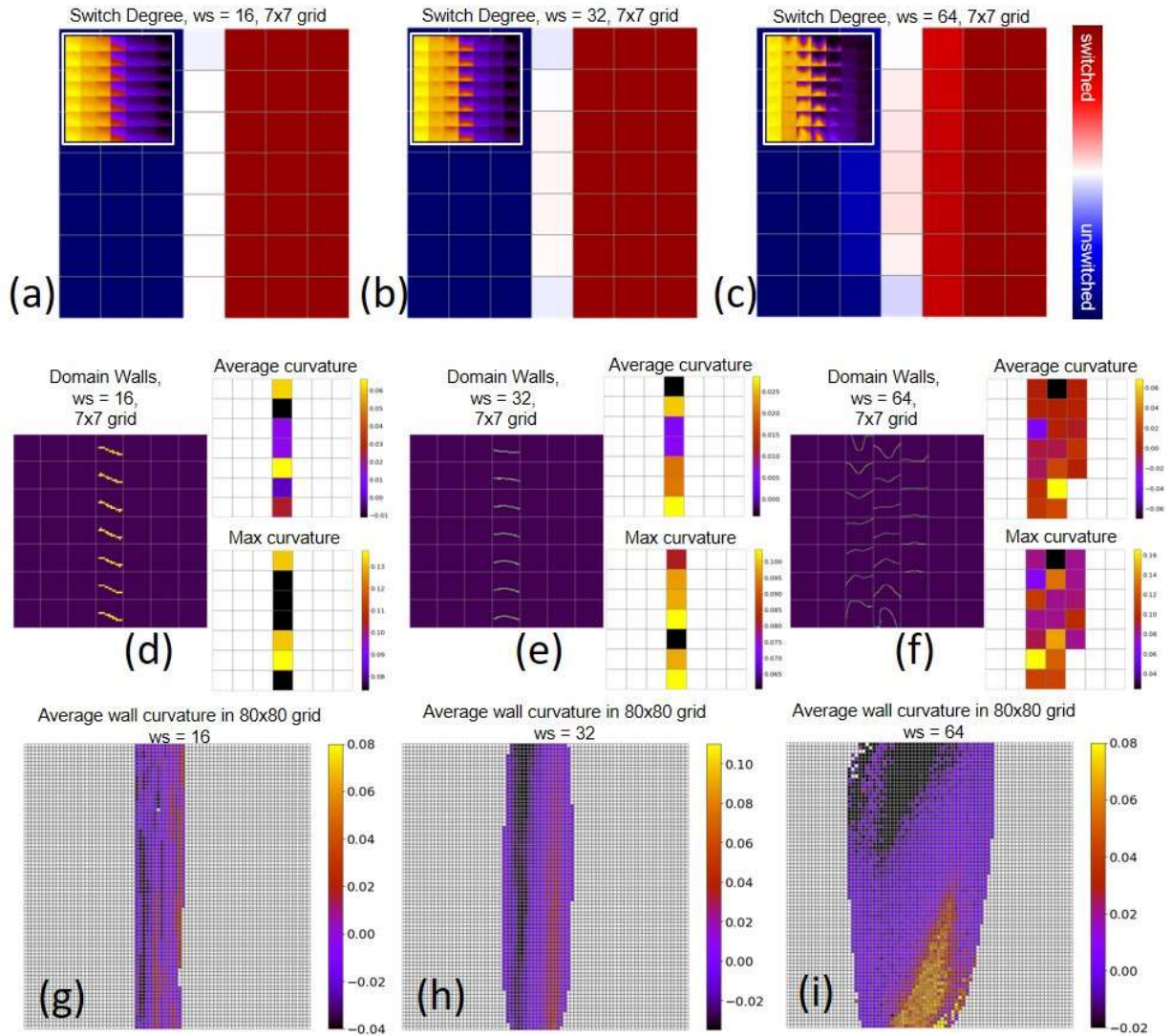

**Figure 4.** The relationship between latent space variables and physics-based descriptor. (a-c) Degree of switching corresponding to latent spaces of rVAE models trained on different window sizes; insets show the corresponding latent space manifolds. (d-f) Domain wall curvature corresponding to the latent manifolds in the insets of (a-c), the average curvature and the maximum curvature for the grid columns with detected domain walls. (g-i) Average domain wall curvature corresponding to 80*80 latent space grid.

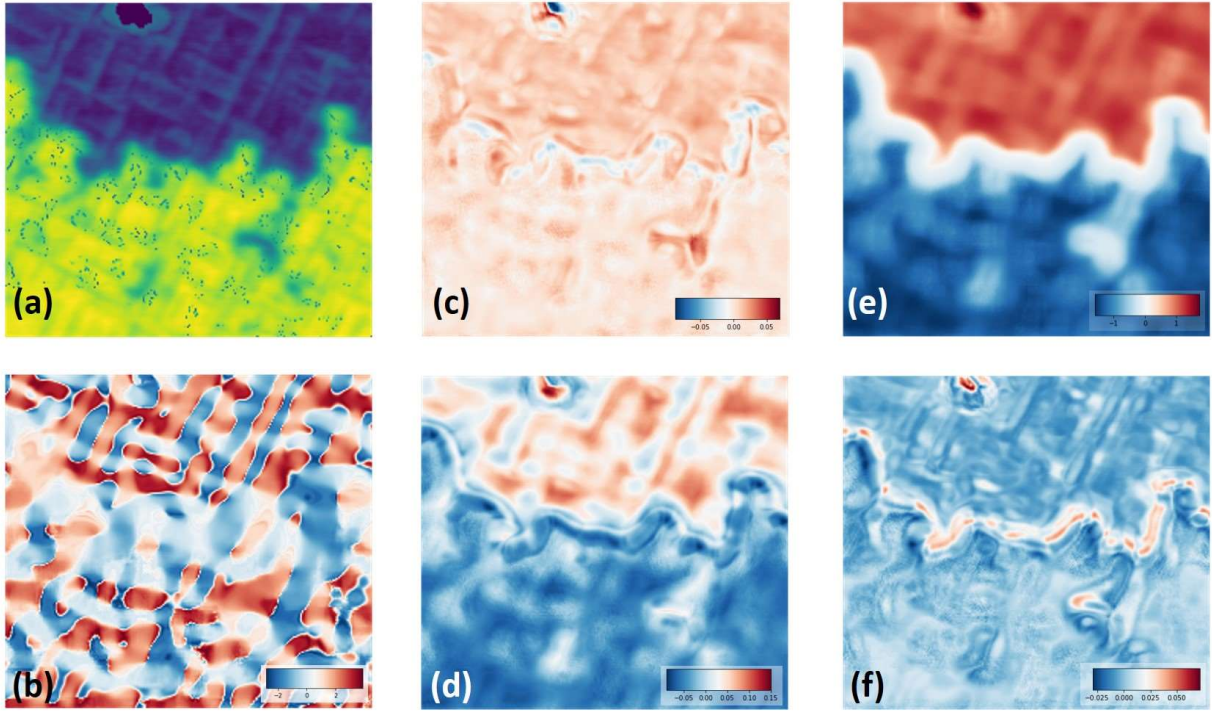

**Figure 5.** Latent variable maps for *n* = 16. (a) Original PFM image (similar to Fig. 1 (c)). (b) Angle, (c) x offset, (d) y offset, (e) latent 1 and (f) latent 2 maps. All images are 1.5 μm.

The rVAE analysis can further be extended to encode the original PFM images as shown in Figure 5. Here, Figure 5 (a) shows one original PFM image from the overall switching dataset with a positive tip bias as in Figure 1(a-d). This features a 180 degree domain wall front propagating from bottom to top, with the more subtle *a-c* domain wall pattern also present. The calculated angle map, Fig. 5 (b), can be readily interpreted as an angle towards the nearest domain wall for every point in (a). Note the characteristic cross-hatch pattern emerging due to the *a-c* wall system. The offset maps shown in Fig. 5 (c,d) show relatively weak contrast. This contrast originates due to the fact that the rVAE can "lock" on the wall segment only if the wall is sufficiently close to the center of each window considered. If the wall is far from the sub-image center, the latent variable will correspond to the region in the latent space that does not contain a domain wall (e.g. top or bottom in Fig. 3 (a)), and the offset variable approaches zero. Finally, the latent variable maps are shown in Figures 5 (e, f). Here, the first (non-degenerate) latent variable shows a clear contrast at the domain wall, which is now easily identifiable. At the same time, the second (degenerate) latent variable shows only weak variation across the image but a predominantly high contrast at domain walls, implying that the second latent variable corresponds to the behavior of domain wall. Note that this reconstruction of the image is accomplished in a fully unsupervised manner, an extremely important consideration in the analysis of large volumes of experimental data and for enabling automated experimentation with PFM and other microscopy methods.

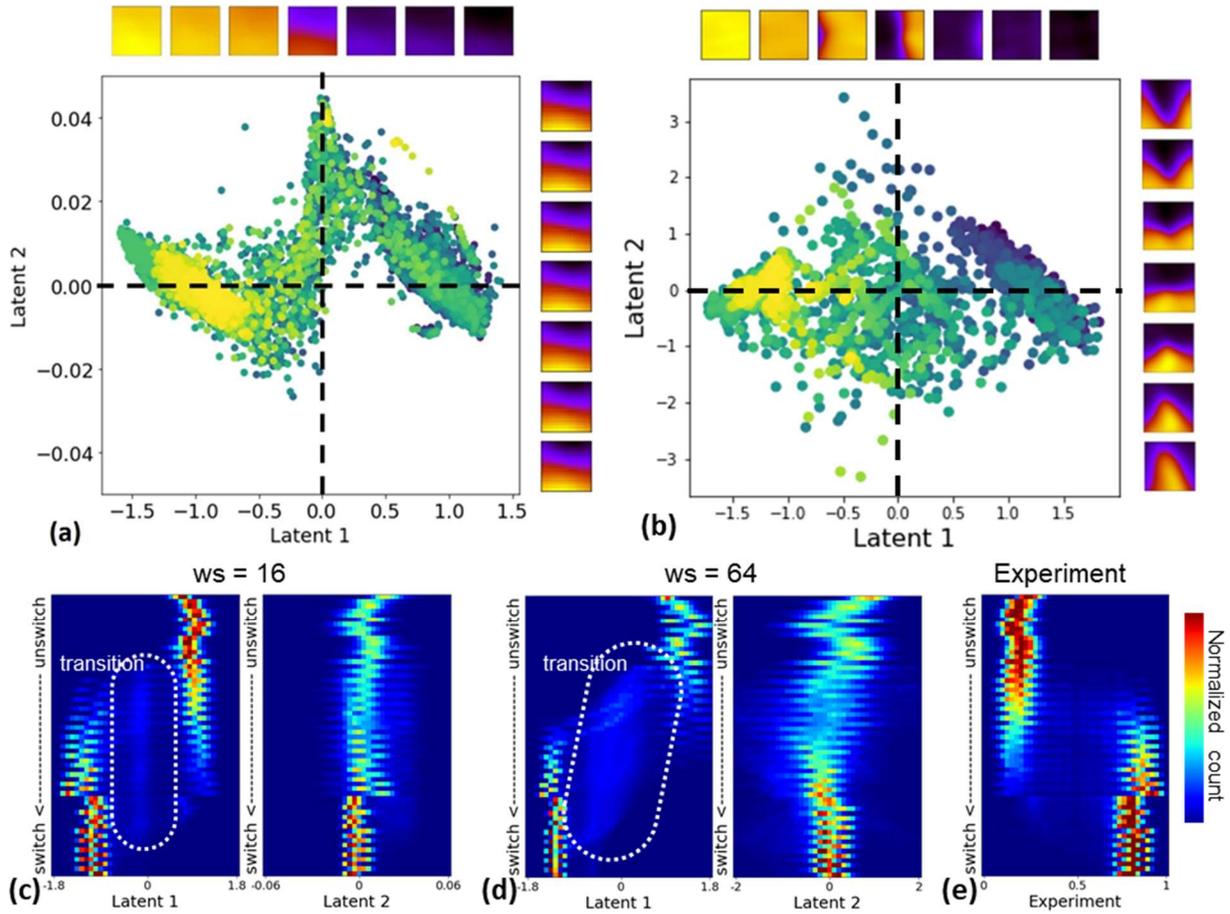

**Figure 6.** Evolution of the domain structure in the latent space during switching. Shown is the evolution of the latent space distributions for the wall-motion scenario for (a) $n = 16$ and (b) $n = 64$. The domain structures reconstructed from latent space correspond to the dotted lines. Note the scale for both latent variables in (a) and (b). (c, d) The evolution of latent variables during the domain switching for window size of 16 (c) and window size 64 (d). (e) The evolution of the experimental data. Note that the first latent variable shows a transition state for both window sizes (c-d), while this transition state cannot be directly observed from experiment data (e).

Finally, this approach allows to explore the ferroelectric switching pathways via analysis of the latent space variables. Figure 6 presents the evolution of the domain structure in the latent space for the domain wall motion scenario (top row in Figure 1). Figure 6 (a) shows the behavior for the small window size, $n = 16$ (incorporating an area of almost 100x100 nm), in which case rVAE acts as the edge filter. To interpret the data in terms of the characteristic domain configurations in real-space, the dashed lines represent the domain geometries. The transition from the unswitched (dark, on the right) domains towards fully switched state (bright, on the left) is clearly seen, and though consistent the magnitude is notably weak along the second latent variable axis (the ordinate). This corresponds to the degeneracy of latent space in this direction. In other words, at this length scale the domain wall is essentially 1D. Despite this, the trajectory indicates a clear zig-zag shape that can be interpreted as systematic change in wall morphology during switching.

In comparison, the analysis for the *n* = 64 window is shown in Fig. 6 (b). In this case, the domain structure changes considerably both along the first and second latent direction. Here, the first latent variable best describes switching from the down to up states, whereas the second latent variable is now a descriptor for the domain wall shape and can be associated with wall curvatures in Fig 3 (f). For intermediate switching steps, a broad spectrum of wall geometries is identifiable. Analyzing histograms of the evolution of latent variables during domain switching provides more physical insights into the switching dynamics. Figures 6c,d show the evolution of latent variables for window sizes of 16 and 64, respectively. The first latent variable exhibits a clear transition state during the domain switching, as marked in Figures 6c,d. This transition state is clearer for the larger window size of 64 (Figure 6d). Meanwhile, such a transition state cannot be directly observed from experimental data, as illustrated in Figure 6e. This suggests that rVAE-based analyses can potentially offer a new understanding of physical mechanisms. In addition, the second latent variable shows multiple abrupt changes at the beginning of the switching process suggesting a domain wall is unstable when initially formed (undergoing changes under the application of electric field). At the same time, the second latent variable is relatively stable at the end of switching process, indicating a stabilization of the domain wall. This behavior confirms that domain wall dynamics are critical to domain switching in ferroelectric materials, that is, at the beginning of switching, it potentially plays a role in facilitating domain switching and then stabilizes when the switching process approaches completion.

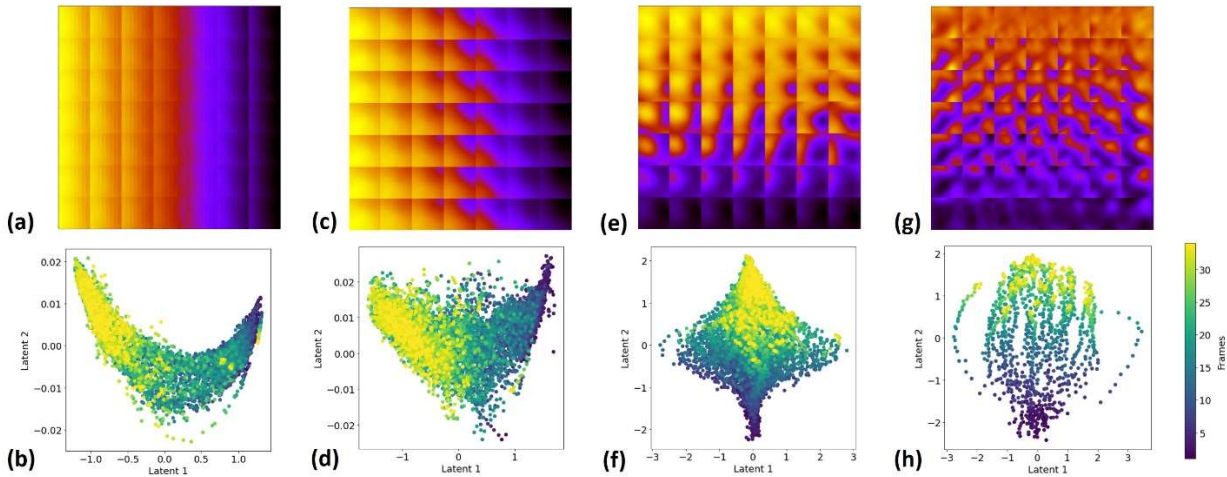

**Figure 7.** Evolution of the domain structure in the nucleation-dominated scenario. Shown is the latent space (top row) and evolution (bottom row) for (a,b) *n* = 8, (c,d) *n* = 16, (e,f) *n* = 32 and (g,h) *n* = 64. Note that the selection of the latent variables is random (compare orientations for a,c,e,g); however, the complexity and degree of switching are disentangled in all cases.

Figure 7 similarly analyzes the evolution of the domain structure during polarization, but for the scenario dominated by domain nucleation instead of growth (Fig. 1e-f). This analysis is performed for several window sizes, exploring progressively more complex details of domain wall structure. For small windows, $n = 8$ and 16 (corresponding to ~50 and 100 nm respectively), the latent space is degenerate, i.e. strong variability is observed only in one direction (latent 1, the abscissa) just as occurred with the results in Figure 5 (a). For these conditions, rVAE is again effectively an adaptable edge filter, as seen in Fig. 6 (a, c). The evolution in the latent space correspondingly represent the simple change of areal fraction from one domain orientation to another.

For larger window sizes, however, more complex behavior is observed. For $n = 32$ the latent space is no longer degenerate (Fig. 7 (e)), and the latent variables now disentangle the switching (ordinate) and domain wall complexity (abscissa). The evolution of the system now can be represented as a gradual transition from the down polarized state with low complexity, through an intermediate polarized state with high complexity, and concluding with the up polarized state with low complexity again. Finally, for $n = 64$, the latent space variables now relate to the domain shape, since the window size becomes comparable to the domain size of the numerous nucleating domains. The latent space in Fig. 7 (h) consequently distinguishes the dynamics of the 20-30 individual domains while they switch.

To summarize, here we introduce a machine learning method for the analysis of the ferroelectric domain switching mechanisms in PFM based on rotationally-invariant variational autoencoders. The rVAE approach allows a simplification of the elements of the observed domain structure, encapsulating common spatial offsets and rotational invariance while reducing the variability of local polarization distributions to a small number of latent variables. For small window sizes the latent space is degenerate, with appreciable variability observed only in the direction of a single latent variable. In this case, the rVAE effectively becomes an edge filter allowing for arbitrary rotations, and analysis of the latent space distributions yields insight into wall angle distributions. This is important in and of itself, since such angular responses can couple to underlying features or processes when studying dynamics—in this case the domain microstructure within the thin ferroelectric film. Of course, such an analysis is feasible with less sophisticated routines, but for larger window sizes the calculated latent space becomes demonstrably 2D. The latent variables then disentangle the degree of switching from the complexity of the domain structure. The switching pathways of the system can thus be simply and efficiently visualized in latent space, now providing unique insight into domain evolution mechanisms. In particular, this allows us to discover a transition state during switching from observing the evolution of latent variables. The rVAE's latent variables also provide insights into the domain wall behavior, such as the wall dynamics and changes in the curvature. In addition, the evolution of latent variables clearly reveals domain wall dynamics at different switching stages, connecting the latent space representation with the role of domain walls in polarization switching of ferroelectric materials. In the future, employing rVAE over much larger areas, or even additional dimensions such as voltage or temperature, can efficiently provide important new insights into domain switching behavior.

Finally, this approach is generalizable and not limited to dynamic PFM data of domain walls as considered here merely as a model system. The rotationally-invariant representation or rVAE, combined with the well-known tendency of classical VAEs to yield disentangled representations and parsimony of latent space descriptors, make it an ideal tool for analyzing imaging data in areas such as dynamic electron, optical, and x-ray microscopy, biological imaging, etc. We demonstrate that exploration of the latent space dimensionality, as a function of window size, will provide particular insight into the spatial complexity of a such systems. Similarly, VAEs universally allow a broad set of tunability via the choice of the loss function and potentially the priors in the latent space, suggesting enormous future potential for applying our method to physical and biological imaging of dynamic phenomena.

**Acknowledgements:**

This research was conducted at the Center for Nanophase Materials Sciences, which also provided support (SVK, MZ) and is a US DOE Office of Science User Facility. Y.L. is supported by the Center for 3D Ferroelectric Microelectronics at Pennsylvania State University. J.J.S. and B.D.H. recognize support from the NSF (MRI development award, DMR-1726862). The authors gratefully acknowledge R. Ramesh (UC Berkeley) for the materials used in this study.

**Data and code availability:**

The interactive Jupyter notebook that reproduces this paper's results is available at https://git.io/JJO2P.